# MOSFET Characterization and Modeling at Cryogenic Temperatures

Chao Luo,[1,2] Zhen Li,[1,2] Teng-Teng Lu,[1,2] Jun Xu,[2] Guo-Ping Guo[1,a)]

[1] *Key Laboratory of Quantum Information, University of Science and Technology of China, Hefei, Anhui 230026, China*

[2] *Department of Physics, University of Science and Technology of China, Hefei, Anhui 230026, China*

**Abstract**—Cryogenic CMOS technology (cryo-CMOS) offers a scalable solution for quantum device interface fabrication. Several previous works have studied the characterization of CMOS technology at cryogenic temperatures for various process nodes. However, CMOS characteristics for various width/length (W/L) ratios and under different bias conditions still require further research. In addition, no previous works have produced an integrated modeling process for cryo-CMOS technology. In this paper, the results of characterization of Semiconductor Manufacturing International Corporation (SMIC) 0.18 μm CMOS technology at cryogenic temperatures (varying from 300 K to 4.2 K) are presented. Measurements of thin- and thick-oxide NMOS and PMOS devices with different W/L ratios are taken under four distinct bias conditions and at different temperatures. The temperature-dependent parameters are revised and an advanced CMOS model is proposed based on BSIM3v3 at the liquid nitrogen temperature (LNT). The proposed model ensures precision at the LNT and is valid for use in an industrial tape-out process. The proposed method presents a calibration approach for BSIM3v3 that is available at different temperature intervals.

## Ⅰ INTRODUCTION

Quantum computers take advantage of two fundamental properties of quantum mechanics, i.e., superposition and quantum entanglement, and are based on qubits. A minimum of a few thousand and possibly millions of qubits are required for quantum computers and these qubits operate at deep cryogenic temperatures. High-precision control electronics are required on a massive scale for the manipulation and readout of single qubits in the so-called fault-tolerant mode[1]. A quantum computer comprises a quantum processor, in which the computations are performed, and a classical-type electronic controller, which is implemented using room-temperature (RT) laboratory instrumentation. The wiring requirements between the cryogenic quantum processor and the RT read-out controller mean that the electronic interface required for control of the quantum processor can be both expensive and unreliable. As an alternative, an electronic interface operating at cryogenic temperatures has been proposed. Cryo-CMOS enables scalability in quantum processors that will allow improved system compactness and cryogenic operation, but a few challenges, including the power limitations of the refrigerators, the interconnections, the packaging and device modeling, have still to be faced[2].

Therefore, CMOS modeling is a priority subject for cryogenic studies.

BSIM3v3 is an industry-standard model that is valid over the temperature range from 230 K to 430 K. However, the characteristics of metal-oxide-semiconductor field-effect transistors (MOSFETs) change because of the freeze-out effect, which has stimulated a requirement for a suitable SPICE model to be developed for use at cryogenic temperatures. Cryogenic temperature operation of MOSFETs offers advantages that include improved turn-on capabilities, high gain, latch-up immunity, improved driving capabilities, leakage current reduction, lower thermal noise, higher integration density, and low power consumption.

CMOS technologies ranging from the 3 μm node down to 14 nm have been characterized over the temperature range from 100 K down to 4.2 K [1,3-16]. A brief review, which is illustrated in Fig. 1, compares the results of this work with those of the previous publications.

In this paper, characterization of Semiconductor Manufacturing International Corporation (SMIC) 0.18 μm CMOS transistors is presented at various temperatures over the range from 300 K to 4.2 K. In addition, an advanced CMOS model based on BSIM3v3 is proposed for use at the LNT. Measurements are performed using the B1500A semiconductor device analyzer and modeling is performed using BSIMProPlus. The temperature-dependent parameters are revised and the model shows good agreement with the measurement results. The method provides a calibration approach for the BSIM3v3 parameters that is available for a variety of different process nodes and temperature intervals. A comparison with previous works on integrated modeling under various W/L and bias conditions is also provided. The model presented is well matched with the industrial process. The research described here also has good application prospects in fields including space exploration, cryogenic electronics and quantum computing. However, modeling at the liquid helium temperature (LHT) will require further research.

## II CRYOGENIC MEASUREMENTS

Measurements of the CMOS transistors were performed using both the thin (3.87 nm) and thick oxide (11.9 nm) SMIC 0.18 μm technologies and for a wide range of feature sizes, as shown in Table I. All the electrical measurements were performed using the Agilent B1500A semiconductor device analyzer. Measurements were taken at different cryogenic temperatures ranging from 298 K to 4.2 K. For the thin-oxide N-type MOS (NMOS) devices, drain-source current ($I_{DS}$) versus drain-source voltage ($V_{DS}$) curves were measured under two substrate bias voltages ($V_{BS}$=0 V and −1.8 V), while sweeping $V_{DS}$ from 0 to 1.8 V in 0.05 V steps, and for values of $V_{GS}$ =0 V to 1.8 V in 0.3 V steps; $I_{DS}$ versus gate-source voltage ($V_{GS}$) curves were also measured for two drain-source bias values ($V_{DS}$=0.05 V and 1.8 V), while sweeping $V_{GS}$ from 0 to 1.8 V in 0.05 V steps, and for values of $V_{BS}$=0 V to −1.8 V in 0.3 V steps. For the thick-oxide NMOS/P-type MOS (PMOS) devices, the measurement bias conditions were similar to those used for the thin-oxide devices, as indicated in the captions of Fig. 3, Fig. 4 and Fig. 5.

The drain current ($I_D$) increases with decreasing temperature (Fig. 2) because of the increased carrier mobility. An obvious kink is shown for a large $V_{DS}$. However, no kinks show up around 77 K for the maximum operating value of $V_{DS}$. The threshold voltage ($V_{th}$) increases at cryogenic temperatures for both NMOS and PMOS devices, as shown in Fig. 6(a). As shown in Fig. 4(c) and (d), the threshold slope (SS) at LHT for both the thin- and thick-oxide NMOS devices is steeper with respect to the RT. The SS of the thick-oxide device is 1.51 times steeper at the LNT and 3.43 times steeper at the LHT with respect to RT. In addition, the transconductance ($G_m$) tends to increase in proportion to decreasing $T$ (Fig. 6(c)).

**Table I SUMMARY OF CHARACTERIZED DEVICES**

| Device Size | Thin-oxide NMOS/PMOS | | Thick-oxide NMOS/PMOS | |
|---|---|---|---|---|
| | W(μm) | L(μm) | W(μm) | L(μm) |
| SMIC 0.18 μm | 100 | 0.18 | 100 | 0.6/0.5 |
| | 10 | 10 | 10 | 10 |
| | 10 | 0.6 | 10 | 0.65/0.55 |
| | 10 | 0.2 | 10 | 0.5/0.45 |
| | 10 | 0.18 | 10 | 0.2 |
| | 10 | 0.16 | 0.3 | 0.6/0.5 |

## Ⅲ DISCUSSION ON CRYOGENIC BEHAVIORS

The most noticeable irregularity in the $I_{DS}$-$V_{DS}$ characteristics (Fig. 2(b)) is the kink that occurs in the near mid-$V_{DS}$ range at 4.2 K. This phenomenon is ascribed to the self-polarization of the bulk at cryogenic temperatures[17]. The high electric field results in the presence of impact-ionized carriers along the channel. The self-polarization leads to a significant bulk current that raises the internal body potential and causes a reduction in the apparent $V_{th}$ above a certain $V_{DS}$ value. Therefore, a jump occurs in $I_D$. This irregular behavior can be modeled using a double $V_{th}$ model[18], where $I_D$ is given by the following formula.

$$I_D = K_P \frac{W}{L}(V_{GS} - V_T)^2 \quad (1)$$

with

$$V_T = V_{TMAX} \ for \ V_{DS} < V_{DS*} \ and$$

$$V_T = V_{TMIN} \ for \ V_{DS} > V_{DS*}. \quad (2)$$

where $V_{DS*}$ is the critical value for kink effect.

No kink occurs around 77K for the maximum operating value of $V_{DS}$. The deviation between BSIM3v3 model and characteristics at LNT is mainly ascribed to the freeze-out effect of source-drain parasitic resistance[19]. In CMOS process node, lightly doped drain (LDD) structure is employed to reduce the hot carrier effect and short channel effect. The ionization of impurities will decrease as temperature decreases, especially in lightly doping regions near source and drain. This results in a large series resistor in the LDD region at low temperatures. When the source-drain voltage becomes very high, the ionization of impurities will be activated to restrain the freeze-out effect under strong electric field. In summary, MOSFET characteristics are manifested as suppressed leakage current in deep linear regions and normal-state at high source-drain voltages. It should be pointed out that the performances of devices with higher doping are not typically degraded by freeze-out, as was previously the case for lightly doped devices[20].

When the temperature decreases, $V_{th}$ increases as expected, as shown in Fig. 6(a). The mobility is strongly enhanced because of the reduction in carrier scattering due to lattice vibrations[21,22]. The effective mobility ($\mu_{eff}$) is dependent on the threshold voltage and the gate and substrate voltages. At a low gate voltage, the Coulomb scattering is dominant and $\mu_{eff}$ increases with increasing gate voltage, which then results in an increase in the transconductance. At higher gate voltages, strong electric field effects play a major role, leading to a reduction in the mobility and a corresponding reduction in the transconductance value. Figure 6(b) illustrates this behavior for a thick-oxide NMOS device at various temperatures. The SS is an important parameter for MOSFETs operating as digital logic switches in the sub-threshold region. SS is given by

$$SS(T) = \left[\frac{\partial \log(I_D)}{\partial V_{GS}}\right]^{-1} = \ln(10)\frac{nkT}{q} \quad (3)$$

where k is Boltzmann constant, q the electron charge, T the Kelvin temperature and n is the sub-threshold slope factor.

Cryogenic performance at RT, LNT and LHT are listed and compared in Table II. The parameters that are known to affect the performance of analog circuits strongly all show improvements. An increase in $G_m$ will result in a wider bandwidth being available for the same power budget. Additionally, the concomitant increase in $G_{ds}$ leads to a reduction of the transistor's intrinsic DC gain. Any increase in SS is beneficial for the digital logic performance, leading to increased speed in the sub-threshold region.

Table II COMPARISON OF PERFORMANCE AT RT, LNT AND LHT

| Technology | NMOS,W/L=10μm/10μm | | |
|---|---|---|---|
| Temperature | 4.2K | 77K | 298K |
| $V_{th}$ [V] | 1.164 | 1.153 | 0.799 |
| SS [mV/dec] | 26.14 | 59.31 | 89.86 |
| $G_m(V_{GS}=1.5V)$[mS] | 0.114 | 0.071 | 0.011 |
| $G_{ds}(V_{DS}=1.5V)$[mS] | 0.623 | 0.600 | 0.225 |

## Ⅳ MODELING

Several prior works modified their developed models by adjusting the temperature-dependent parameters based on the results from the LNT measurements and augmented the sub-circuit model using additional electrical components[23]. A cryogenic modeling methodology is presented here that allows temperature-dependent parameters to be adjusted by inserting a correction coefficient into the BSIM equations and then solving the modified equations[24]. Another parameter extraction method for use at low temperatures was introduced elsewhere[25].

The basic idea consisted of construction of a function that can eliminate the effects of the mobility's gate voltage dependence.

In this work, the extraction procedure is performed using BSIMProPlus. Before the temperature-dependent parameter extraction process, the temperature of the simulator must be set to 77 K to match the SS. The parameters with the greatest influence on $I_D$ are extracted first, i.e., $V_T$ (vt0, K1, K2, dvt0 for BSIM) and $\mu$ (μ0, μa, μb, μc). Additional modeling was performed by taking the drain/source charge sharing behavior (dvt0, dvt1, dvt2), the carrier drift velocity (vsat) and the bulk charge effect (a0, ags, keta) into consideration. The remaining parameters would lead to further enhancement of the fitting but have only a minor influence on the device characteristics.

The root-mean-square (RMS) error is then introduced to estimate the deviation between the results from the measurements and the simulations. The RMS is given by

$$\text{RMSerr} = \sqrt{\frac{1}{N}\sum_{i=1}^{n}\left(\frac{I_{measi}-I_{simui}}{I_{threshold}}\right)^2} \times 100 \quad (4)$$

where N is the number of data, $I_{measi}$ is measured data and $I_{simui}$ is simulated data. The value of the threshold current $I_{threshold}$ can be set appropriately to obtain meaningful results. In this case, $I_{threshold}$ has been set to the maximum measured value according to BSIMProPlus.

**Table III RMS ERROR of MEDOL WITH DEFAULT AND REVISED PARAMETERS.**

| (Thick-oxide) | NMOS | | PMOS | |
|---|---|---|---|---|
| W/L=10μm/10μm | default | revised | default | revised |
| $I_D(V_{GS})(V_{DS}=0.1V)$ | 35.31% | 1.53% | 20.07% | 1.63% |
| $I_D(V_{DS})(V_{BS}=0V)$ | 33.29% | 1.57% | 20.97% | 1.36% |

## Ⅴ SIMULATION VERSUS MEASUREMENT

Good agreement with the DC measurements was achieved for devices operating at the LNT. Measurements and simulations of the $I_{DS}$-$V_{DS}$ and $I_{DS}$-$V_{GS}$ curves for the NMOS (Fig. 7(a)–(f)) and PMOS (Fig. 7(g)–(h)) devices are plotted at the LNT. The RMS errors were calculated for the measured data with respect to the simulations using both the default and revised parameters, as summarized in Table III. A significant optimization of the BSIM3v3 model has been achieved.

## VI CONCLUSION

A study of the performance of SMIC 0.18 μm/1.8 V CMOS technology at cryogenic temperatures has been presented in this work. An advanced model based on BSIM3v3 has been proposed to optimize the deviations between the measurement results and the results of simulations using the default parameters. The proposed method presents an optimization approach using temperature-dependent parameters that can be extended to any temperature region. Improvements in the proposed model will allow precise description of MOSFETs to aid in the design of more complex cryogenic circuits and systems. Further research into modeling at the LHT will continue as part of our future work.

## VII ACKNOWLEDGMENT

The authors would like to thank SMIC for devices fabrication and software support. This work was supported by the National Key Research and Development Program of China (Grant No.2016YFA0301700), the National Natural Science Foundation of China (Grants No. 11625419), the Anhui initiative in Quantum information Technologies (Grants No. AHY080000) and this work was partially carried out at the USTC Center for Micro and Nanoscale Research and Fabrication.

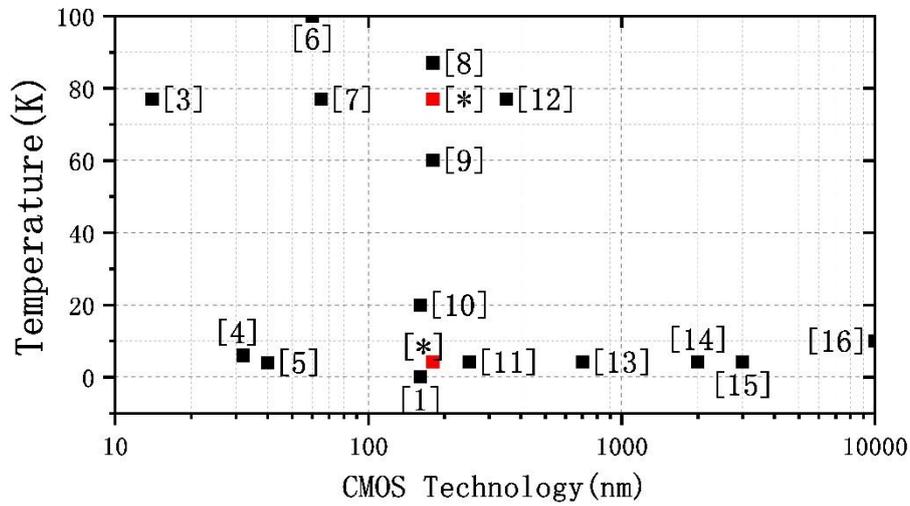

Fig. 1 A brief review of characterization of CMOS transistors at cryogenic temperatures. Prior works are indicated with black dot and devices presented in this paper with red dots.

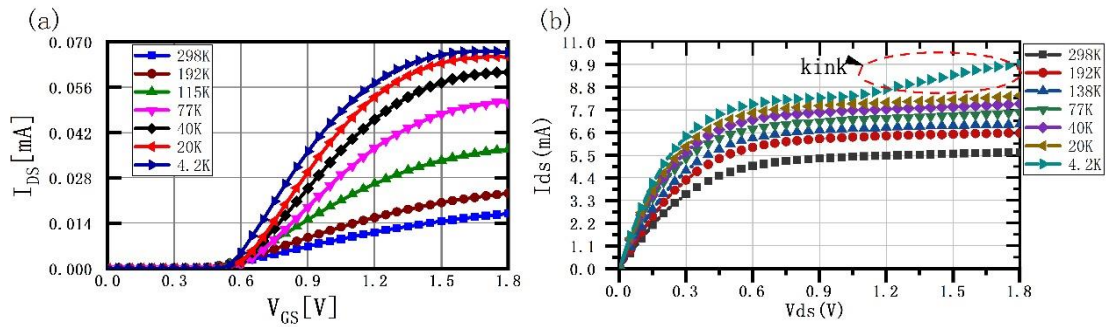

Fig. 2 $I_{DS}$-$V_{GS}$ and $I_{DS}$-$V_{DS}$ of thin-oxide NMOS at different temperatures varying from 298K to 4.2K. W/L in μm. (a) W/L =10/10. (b) W/L =10/0.2.

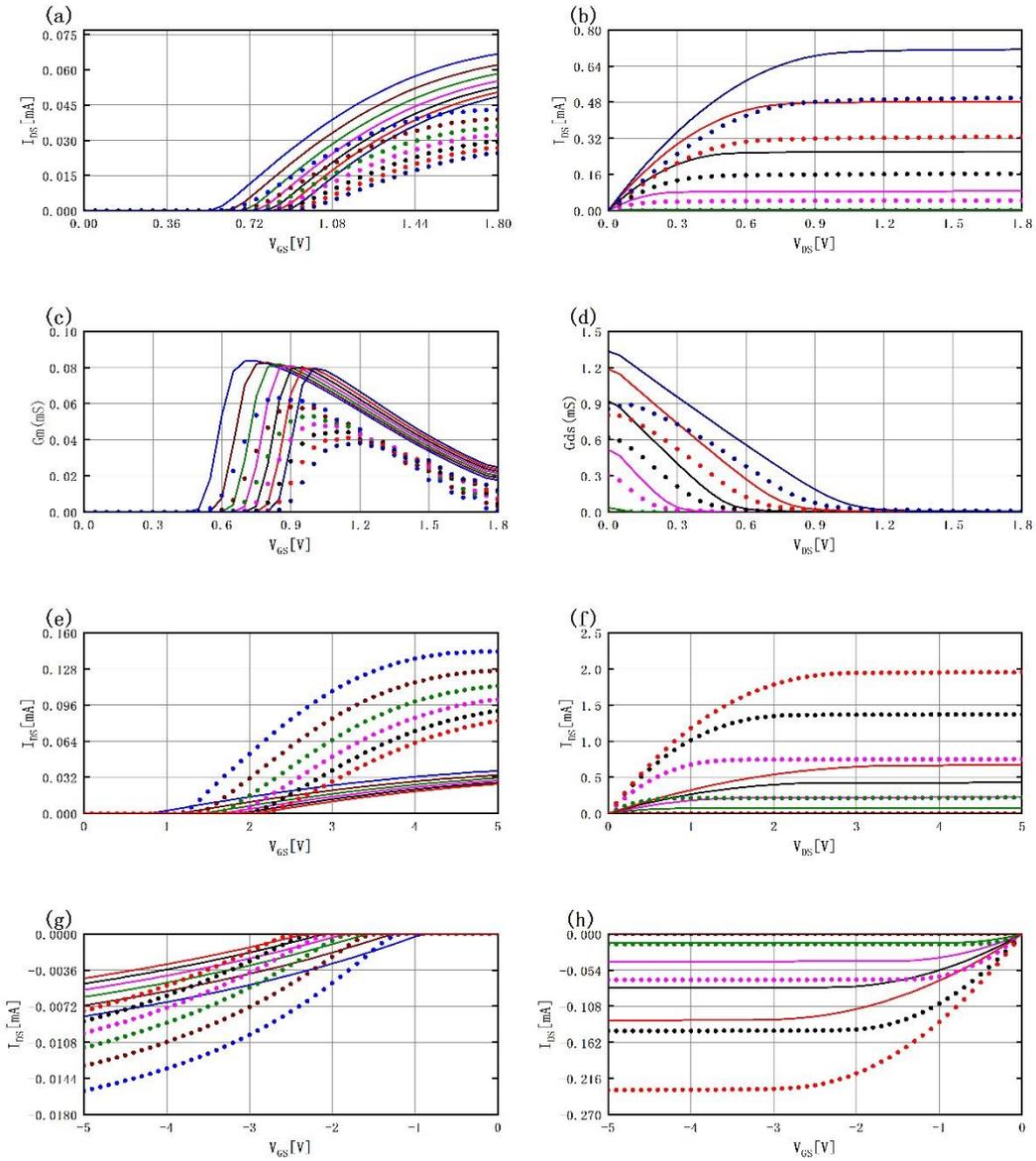

Fig. 3 $I_{DS}$-$V_{GS}$ and $I_{DS}$-$V_{DS}$ of NMOS in thin- (a,b) and thick-oxide (e,f), and thick-oxide PMOS (g,h) at LNT (dashed lines). $G_m$-$V_{GS}$ (c) and $G_{ds}$-$V_{DS}$ (d) of thin-oxide NMOS. The solid lines are simulations of BSIM with non-revised parameters. W/L = 10μm /10μm for all deveices.
(a,c) $V_{DS}$=0.05V, $V_{BS}$= 0 → -1.8V; (b,d) $V_{BS}$=0V, $V_{GS}$= 0 → 1.8V;
(e) $V_{DS}$=0.10V, $V_{BS}$= 0 → -4V; (f) $V_{BS}$=0V, $V_{GS}$= 0 → 5V;
(g) $V_{DS}$=-0.10V, $V_{BS}$= 0 → 4V; (h) $V_{BS}$=0V, $V_{GS}$= 0 → -5V;

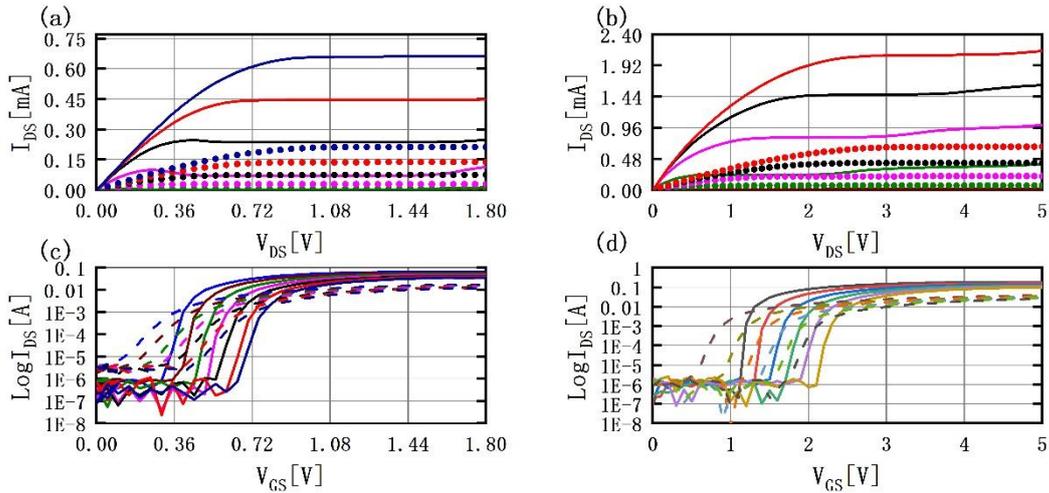

Fig. 4 $I_{DS}$-$V_{DS}$ and $I_{DS}$-$V_{GS}$ of NMOS in thin- (a,c) and thick-oxide (b,d) at RT (dashed lines )and LHT (solid lines). W/L in μm.
(a) W/L=10/10, $V_{BS}$=0V, $V_{GS}$= 0 → 1.8V; (b) W/L=10/10, $V_{BS}$=0V, $V_{GS}$= 0 → 5V;
(c) W/L=10/10, $V_{DS}$=0.05V, $V_{BS}$= 0 → -1.8V; (d) W/L=10/10, $V_{DS}$=0.10V, $V_{BS}$= 0 → -4V.

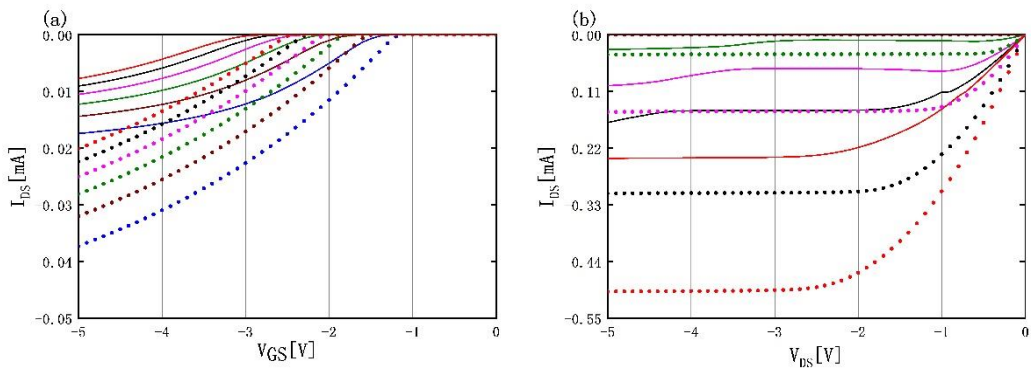

Fig. 5 $I_{DS}$-$V_{GS}$ and $I_{DS}$-$V_{DS}$ measurements of thick-oxide PMOS at RT(dashed lines) and LHT (solid lines). (a) $V_{DS}$=-0.10V, $V_{BS}$= 0 → 4V; (b) $V_{BS}$=0V, $V_{GS}$= 0 → -5V;

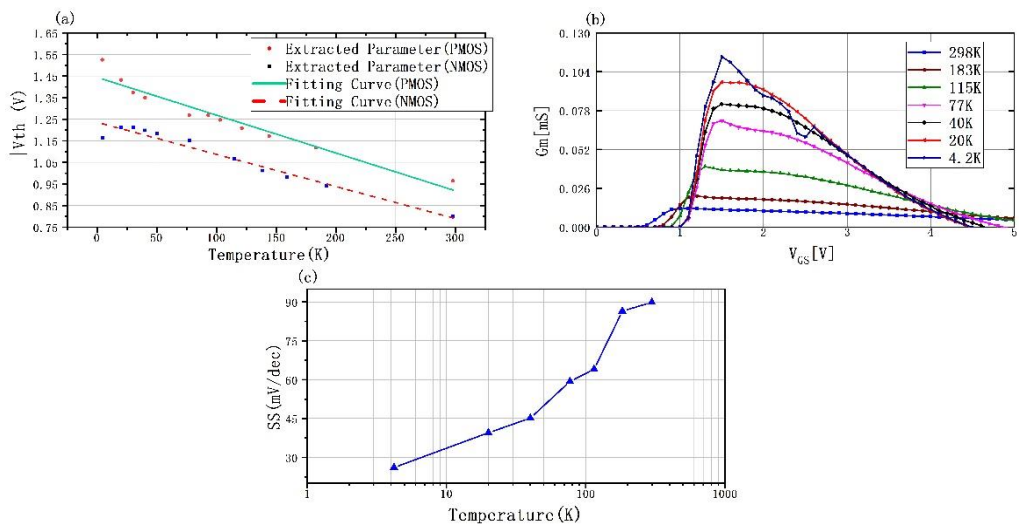

Fig. 6 (a) Variations and fitting curves of extracted threshold voltage at zero substrate bias for different temperatures; (b) Variations of the trans-conductance with gate voltage on a 10μm /10μm thick-oxide NMOS for different temperatures; (c) SS versus temperature of thick-oxide NMOS, W/L=10μm /10μm.

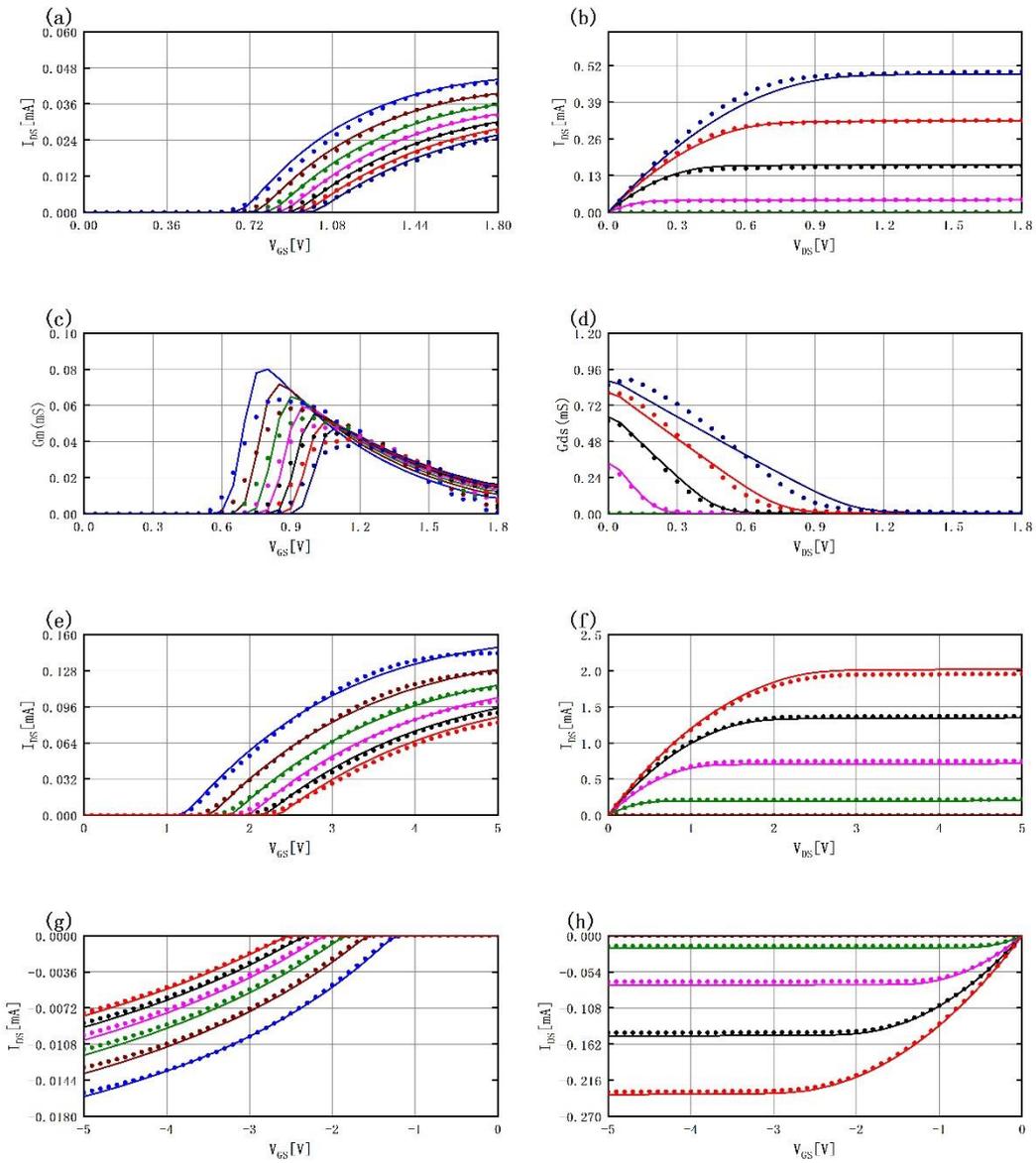

Fig. 7 Revised models (solid lines) and measurements (dashed lines) at LNT. W/L = 10μm /10μm for all deveices.

(a,c) $V_{DS}$=0.05V, $V_{BS}$= 0 → -1.8V; (b,d) $V_{BS}$=0V, $V_{GS}$= 0 → 1.8V;

(e) $V_{DS}$=0.10V, $V_{BS}$= 0 → -4V; (f) $V_{BS}$=0V, $V_{GS}$= 0 → 5V;

(g) $V_{DS}$=-0.10V, $V_{BS}$= 0 → 4V; (h) $V_{BS}$=0V, $V_{GS}$= 0 → -5V;